\documentclass[sigchi]{acmart}

\usepackage{booktabs} 
\usepackage{multirow}
\usepackage[htt]{hyphenat}
\usepackage[export]{adjustbox}
\usepackage{breakcites}
\usepackage{balance}
\usepackage{url}
\usepackage{hyperref}
\usepackage{afterpage}
\usepackage{floatpag}
\usepackage{cleveref}

\crefname{section}{\S}{\S\S}
\Crefname{section}{\S}{\S\S}

\usepackage{xcolor}
\definecolor{OrangeRed}{HTML}{FF4500}
\definecolor{Blue}{HTML}{348ABD}
\definecolor{green}{HTML}{008000}

\newcommand{\ez}[1]{\textcolor{green}{(ez: #1)}}

\copyrightyear{2019} 
\acmYear{2019} 
\setcopyright{acmlicensed}
\acmConference[CHI 2019]{CHI Conference on Human Factors in Computing Systems Proceedings}{May 4--9, 2019}{Glasgow, Scotland UK}
\acmBooktitle{CHI Conference on Human Factors in Computing Systems Proceedings (CHI 2019), May 4--9, 2019, Glasgow, Scotland UK}
\acmPrice{15.00}
\acmDOI{10.1145/3290605.3300892}
\acmISBN{978-1-4503-5970-2/19/05}

\begin{document}
\title{VizNet: Towards A Large-Scale Visualization\\Learning and Benchmarking Repository}




\settopmatter{authorsperrow=5}
\author{Kevin Hu}
\affiliation{
 \institution{MIT Media Lab}
}
\email{kzh@mit.edu}

\author{Snehalkumar `Neil' S. Gaikwad}
 \affiliation{%
 \institution{MIT Media Lab}
}
\email{gaikwad@mit.edu}

\author{Madelon Hulsebos}
\affiliation{%
  \institution{MIT Media Lab}
}
\email{madelonhulsebos@gmail.com}

\author{Michiel A. Bakker}
\affiliation{%
  \institution{MIT Media Lab}
}
\email{bakker@mit.edu}

\author{Emanuel Zgraggen}
\affiliation{%
 \institution{MIT CSAIL}
}
\email{emzg@mit.edu}

\author{C{\'{e}}sar Hidalgo}
\affiliation{%
 \institution{MIT Media Lab}
}
\email{hidalgo@mit.edu}
 
\author{Tim Kraska}
\affiliation{%
  \institution{MIT CSAIL}
}
\email{kraska@mit.edu}
 
\author{Guoliang Li}
\affiliation{%
  \institution{Tsinghua University}
}
\email{liguoliang@tsinghua.edu.cn}

\author{Arvind Satyanarayan}
\affiliation{%
 \institution{MIT CSAIL}
}
\email{arvindsatya@mit.edu}

\author{{\c{C}}a{\u{g}}atay Demiralp}
\affiliation{%
 \institution{MIT CSAIL}
}
\email{cagatay@csail.mit.edu}


\renewcommand{\shortauthors}{K. Hu et al.}

\begin{abstract}
Researchers currently rely on ad hoc datasets to train automated visualization tools and evaluate the effectiveness of visualization designs. These exemplars often lack the characteristics of real-world datasets, and their one-off nature makes it difficult to compare different techniques. In this paper, we present VizNet: a large-scale corpus of over 31 million datasets compiled from open data repositories and online visualization galleries. On average, these datasets comprise 17 records over 3 dimensions and across the corpus, we find 51\% of the dimensions record categorical data, 44\% quantitative, and only 5\% temporal. VizNet provides the necessary common baseline for comparing visualization design techniques, and developing benchmark models and algorithms for automating visual analysis. To demonstrate VizNet's utility as a platform for conducting online crowdsourced experiments at scale, we replicate a prior study assessing the influence of user task and data distribution on visual encoding effectiveness, and extend it by considering an additional task: outlier detection. To contend with running such studies at scale, we demonstrate how a metric of perceptual effectiveness can be learned from experimental results, and show its predictive power across test datasets.

\end{abstract}



\begin{teaserfigure}
  \includegraphics[width=\textwidth]{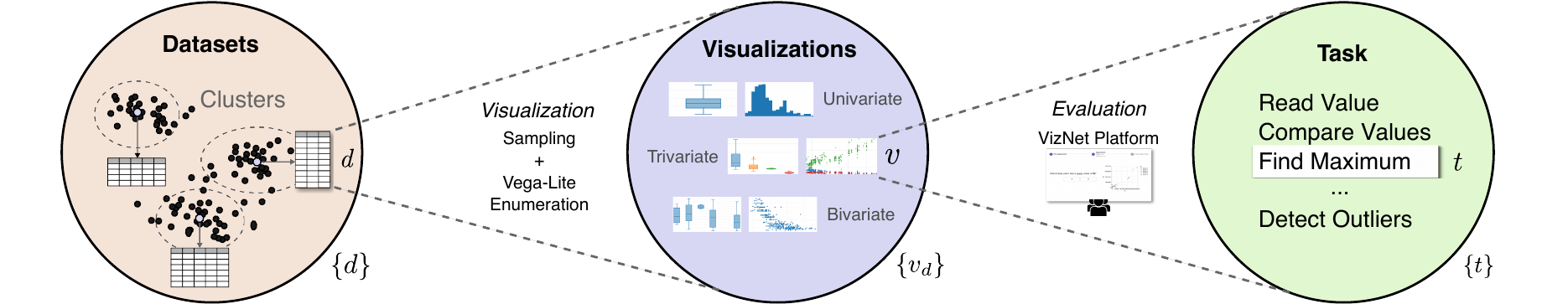}
  \caption{VizNet enables data scientists and visualization researchers to aggregate data, enumerate visual encodings, and crowdsource effectiveness evaluations.}
  \label{fig:teaser}
\end{teaserfigure}

\maketitle
\section{Introduction}

A primary concern in visualization is how to effectively encode data values as visual variables. Beginning with Cleveland and McGill's seminal work~\cite{cleveland-mcgill-graphical-perception}, researchers have studied this question of \emph{graphical perception} by conducting human subjects experiments. And increasingly, researchers are seeking to operationalize the guidelines such studies produce using handcrafted rule-based systems~\cite{2016-compassql, draco} or learned models~\cite{data2vis, deepeye, vizml}. 

To increase the scale and diversity of the subject pool, modern studies have eschewed traditional laboratory setups in favor of crowdsourcing platforms~\cite{heer-and-bostock}. But a constraining factor for true ecological validity remains. Collecting, curating, and cleaning data is a laborious and expensive process and, thus, researchers have relied on running studies with ad hoc datasets. Such datasets, sometimes synthetically generated, do not display the same characteristics as data found in the wild. Moreover, as one-off exemplars, their use makes it difficult to compare approaches against a common baseline. 

Large-scale databases (such as WordNet~\cite{miller1995wordnet} and ImageNet~\cite{Deng09imagenet:a}) have proven instrumental in pushing the state-of-the-art forward as they provide the data needed to train and test machine learning models, as well as a common baseline for evaluation, experimentation, and benchmarking. Their success has led researchers to call for a similar approach to advance data visualization~\cite{eichmann, battle-benchmark}. However, insufficient attention has been paid to design and engineer a centralized and large-scale repository for evaluating the effectiveness of visual designs. 

In response, we introduce VizNet: a corpus of over 31 million datasets (657GB of data) compiled from the web, open data repositories, and online visualization platforms. In characterizing these datasets, we find that they typically consist of 17 records describing 3 dimensions of data. 51\% of the dimensions in the corpus record categorical data, 44\% quantitative, and only 5\% measure temporal information. Such high-level properties, and additional measures such as best statistical fit and entropy, contribute a taxonomy of real-world datasets that can inform assessments of ecological validity of prior studies.

We demonstrate VizNet's viability as a platform for conducting online crowdsourced experiments at scale by replicating the Kim and Heer (2018) study assessing the effect of task and data distribution on the effectiveness of visual encodings~\cite{kim-and-heer}, and extend it with an additional task: outlier detection. While largely in line with the original findings, our results do exhibit several statistically significant differences as a result of our more diverse backing datasets. These differences inform our discussion on how crowdsourced graphical perception studies must adapt to and account for the variation found in organic datasets. VizNet along with data collection and analysis scripts is publicly available at \href{https://viznet.media.mit.edu}{\texttt{https://viznet.media.mit.edu}}.

Data visualization is an inherently combinatorial design problem: a single dataset can be visualized in a multitude of ways, and a single visualization can be suitable for a range of analytic tasks. As the VizNet corpus grows, assessing the effectiveness of these \emph{(data, visualization, task)} triplets, even using crowdsourcing, will quickly become time- and cost-prohibitive. To contend with this scale, we conclude by formulating effectiveness prediction as a machine learning task over these triplets. We demonstrate a proof-of-concept model that predicts the effectiveness of unseen triplets with non-random performance. Our results suggest that machine learning offers a promising method for efficiently annotating VizNet content. VizNet provides an important opportunity to advance our understanding of  graphical perception. 


\section{Related Work}
VizNet is motivated by research in graphical perception, automated visualization based on machine learning, and crowdsourced efforts towards data 
collection for visualization research. VizNet also draws on the digital experimentation capabilities of large-scale machine learning corpora. 

\subsection{Graphical Perception}
Visual encoding of data is central to information visualization. Earlier work has studied how different choices of visual encodings such as position, size, color and shape influence \emph{graphical perception}~\cite{cleveland:jasa84}, the decoding of data presented in graphs. Through human subjects experiments, researchers have investigated the effects of visual encoding on the ability to read and make judgments about data represented in visualizations~\cite{skau2016arcs, cleveland:jasa84, heer:chi10,kong:infovis10, lewandowsky:jasa89, simkin:jasa87, spence:acp91, tremmel:jcgs95}.  Consequently, prior research has provided rankings of visual variables by user performance for nominal, ordinal, and numerical data~\cite{cleveland:jasa84, lewandowsky:jasa89, maceachren:book95,mackinlay:tog86, shortridge:ac82}. Researchers have also studied how design parameters beyond visual encoding variables such as aspect ratio ~\cite{cleveland:book93, heer:infovis06, talbot:infovis11}, size~\cite{heer:chi09,lam:tvcg07,cleveland1982variables}, chart variation~\cite{Talbot_bar, Kosara2016pie}, and axis labeling~\cite{talbot:infovis10} impact the effectiveness of visualizations. Previous studies have evaluated how user task, data types and distributions influence the effectiveness of charts~\cite{saket} and visual encoding variables~\cite{kim-and-heer}. 

Graphical perception experiments in current practice are typically conducted on single datasets with small size and variety, lacking the characteristics of real-world data.  Studies based on ad hoc datasets may provide useful results but are inherently partial and difficult to generalize, reproduce and compare against. VizNet provides a corpus of real-world tables from diverse domains to make it easier for researchers to run visualization design evaluation studies at scale. VizNet is sufficiently rich both in size and variety to satisfy the data needs ocf a substantial number of experimental designs, facilitating the comparison of and reasoning about results from different experiments on a common baseline.

\subsection{Data Collection for Visualization Research}
Although researchers recognize the need for data collection and generation to facilitate evaluation across a broad range of real datasets~\cite{sedlmair2012taxonomy,schulz2016generative}, little effort 
has been made to create centralized corpora for data visualization research. Beagle~\cite{battle-beagle} has been used to scrape over 41,000 visualizations from the web. Similarly, the MassVis~\cite{Borkin:2013:MassVis} database was compiled by scraping over 5,000 visualizations from the web and partially annotating them. Lee et al.~\cite{Lee:2018:Viziometrics} recently extracted and classified 4.8 million figures from articles on PubMed Central. However, these datasets do not include the raw data represented by the visualizations, limiting their utility for generalized and reproducible visualization research. 

\subsection{Automated Visualization using Machine Learning}
Data-driven models based on responses elicited through human subjects experiments are common in the psychophysics and data visualization literature. For example, low-level perceptual models such as the Weber-Fechner Law, Stevens' Power Law, the CIELAB color space, and perceptual kernels~\cite{Demiralp_2014b} all fit various models to empirical user data, informing low-level visual encoding design. Earlier researchers propose using such models to generate and evaluate visualizations (e.g., \cite{Demiralp_2014a, Demiralp_2014b, sedlmair2015data}). 

In a natural extension to these earlier ideas, researchers have recently introduced machine learning-based systems for automated visualization design. Data2Vis~\cite{data2vis} uses a neural machine translation approach to create a sequence-to-sequence model that maps JSON-encoded datasets to Vega-lite visualization specifications. Draco-Learn~\cite{draco} learns trade-offs between constraints in Draco. DeepEye~\cite{deepeye} combines rule-based visualization generation with models trained to  classify a visualization as ``good" or ``bad" and  rank lists of visualizations. VizML~\cite{vizml} uses neural networks to predict visualization design choices from a corpus of one million dataset-visualization pairs harvested from a popular online visualization tool. Results from this recent work are promising but also point at the need for large-scale real-world training data with sufficient diversity~\cite{Saket:2018:Beyond}. VizNet addresses this research gap and provides 31 million real-world datasets from everyday domains and can be used for training machine learning models to drive visualization systems. 

\subsection{Machine Learning Corpora}
Recent developments of large-scale data repositories have been instrumental in fostering machine learning research. Access to rich, voluminous data is crucial for developing successful machine learning models and comparing different approaches on a common baseline. To this end, researchers have created centralized data repositories for training, testing, and benchmarking models across many tasks. Publicly available repositories such as ImageNet~\cite{Deng09imagenet:a}, SUN~\cite{Xiao:2016:SUN}, COCO~\cite{Lin:2014:COCO}, etc. are one of the main drivers behind the rapid advances in deep learning. VizNet is informed and inspired by the digital experimentation capabilities of large-scale data repositories in machine learning research. 


\newcommand\webtablesColMean{5.03}
\newcommand\webtablesColMedian{4}

\newcommand\plotlyColMean{21.06}
\newcommand\plotlyColMedian{3}

\newcommand\manyeyesColMean{5.62}
\newcommand\manyeyesColMedian{2}

\newcommand\opendataColMean{12.24}
\newcommand\opendataColMedian{2}

\newcommand\viznetColMean{10.99}
\newcommand\viznetColMedian{3}

\newcommand\webtablesRowMean{14.22}
\newcommand\webtablesRowMedian{5}

\newcommand\plotlyRowMean{5,911.65}
\newcommand\plotlyRowMedian{50}

\newcommand\manyeyesRowMean{963.94}
\newcommand\manyeyesRowMedian{19}

\newcommand\opendataRowMean{14,252.53}
\newcommand\opendataRowMedian{70}

\newcommand\viznetRowMean{5,258.10}
\newcommand\viznetRowMedian{17}

\newcommand\webtablesCPercent{57.58}
\newcommand\webtablesQPercent{35.56}
\newcommand\webtablesTPercent{6.86}

\newcommand\plotlyCPercent{17.29}
\newcommand\plotlyQPercent{75.47}
\newcommand\plotlyTPercent{7.24}

\newcommand\manyeyesCPercent{51.58}
\newcommand\manyeyesQPercent{46.04}
\newcommand\manyeyesTPercent{2.48}

\newcommand\opendataCPercent{76.55}
\newcommand\opendataQPercent{21.20}
\newcommand\opendataTPercent{2.24}

\newcommand\viznetCPercent{50.71}
\newcommand\viznetQPercent{44.58}
\newcommand\viznetTPercent{4.71}

\newcommand\webtablesQDist{normal, log-normal,\\power law}
\newcommand\plotlyQDist{log-normal, normal,\\power law}
\newcommand\manyeyesQDist{normal, log-normal\\power law}
\newcommand\opendataQDist{normal, log normal,\\exponential}
\newcommand\viznetQDist{normal, log-normal,\\power law}

\newcommand\webtablesCEntropyMean{0.94}
\newcommand\plotlyCEntropyMean{0.68}
\newcommand\manyeyesCEntropyMean{0.81}
\newcommand\opendataCEntropyMean{0.50}
\newcommand\viznetCEntropyMean{0.79}

\section{Data}\label{sec:data}
VizNet incorporates four large-scale corpora, assembled from the web, online visualization tools, and open data portals.

\subsection{Corpora}
The first category of corpora includes data tables harvested from the web. In particular, we use horizontal relational tables from the WebTables 2015 corpus~\cite{webtables}, which extracts structured tables from the Common Crawl. In these tables, entities are represented in rows and attributes in columns.

The second type of corpus includes tabular data uploaded by users of two popular online data visualization and analysis systems. Plotly~\cite{plotly} is a software company that develops visualization tools and libraries. Once created, Plotly charts can be posted to the Plotly Community Feed~\cite{plotly-community-feed}. Using the Plotly API, we collected approximately 2.5 years of public visualizations from the feed, starting from 2015-07-17 and ending at 2018-01-06. The second system, ManyEyes~\cite{viegas-manyeyes} allowed users to create and publish visualizations through a web interface. It was available from 2007--2015, and was used by tens of thousands of users~\cite{morton-public-data-and-visualizations}.

The third type of corpus includes public data from the Open Data Portal Watch~\cite{Neumaier2016, characteristics-of-open-data-csv-files}, which catalogs and monitors 262 open data portals such as \href{https://data.noaa.gov}{\texttt{data.noaa.gov}} from CKAN, \href{https://finances.worldbank.org}{\texttt{finances.worldbank.org}} from Socrata, and \href{https://opendata.brussels.be}{\texttt{opendata.brussels.be}} from OpenDataSoft. The majority of these portals are hosted by governments, and collect civic and social data.

VizNet aggregates these corpora into a centralized repository. However, the majority of datasets are from WebTables. Therefore, in the following sections, we describe each corpus individually with 250K randomly sampled datasets, to avoid oversampling the WebTable corpus. We combine these datasets into a balanced sample of one million datasets, which we refer to as the \textbf{VizNet 1M corpus}.

\subsection{Characterization}

\begin{figure*}[p]
    \includegraphics[width=\textwidth]{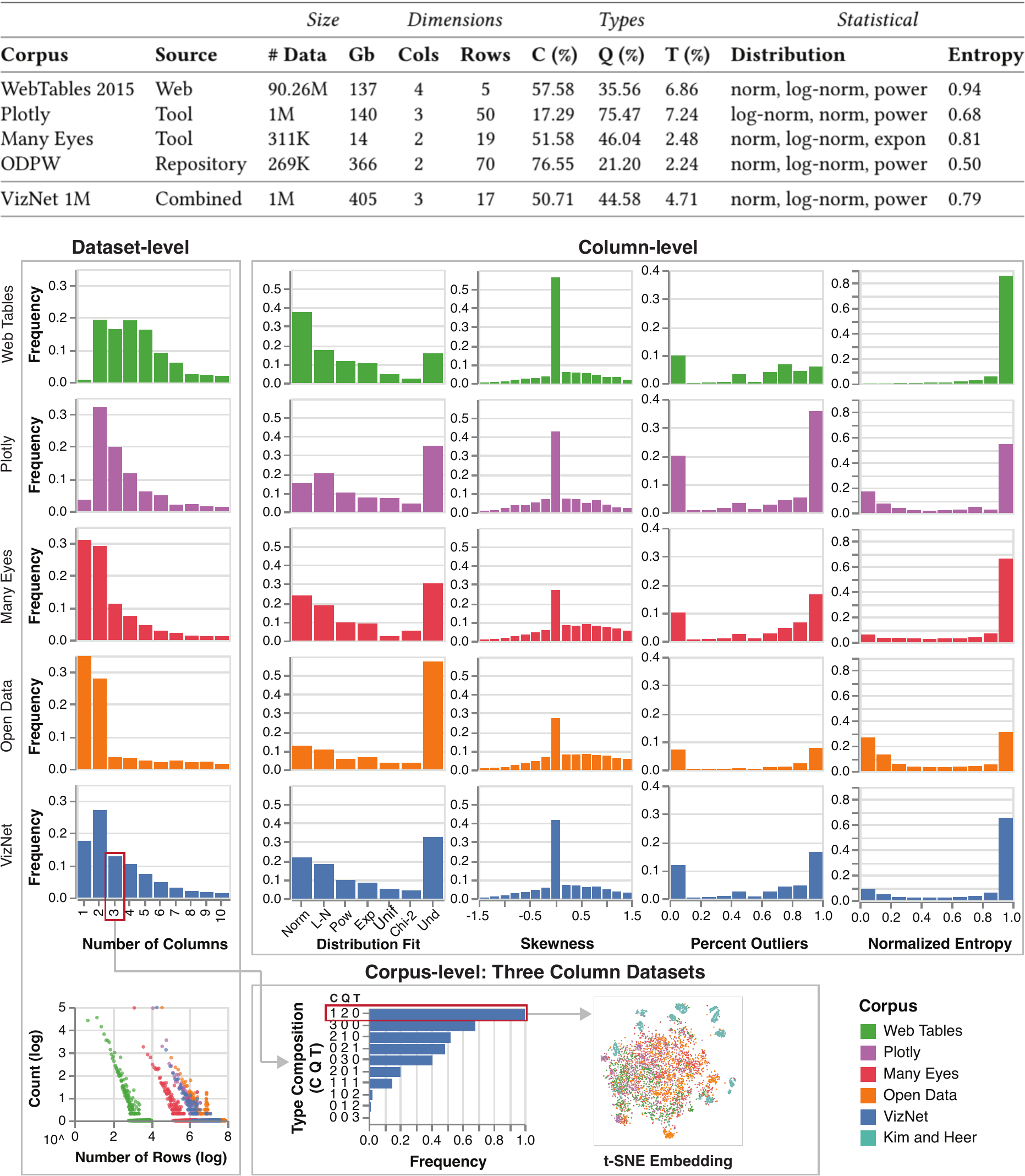}
    \floatpagestyle{empty}
    
    \vspace{-15pt}
    \caption{ Summary statistics and distributions of the four source corpora and the VizNet 1M corpus. Distributions are abbreviated as Norm = normal, L-N = log-normal, Pow = power law, Exp = exponential, Unif = uniform, and Und = undefined. The bars outlined in red represent three column datasets and the subset containing one categorical and two quantitative fields. }
    \label{fig:dataset-one-pager-with-table}
\end{figure*}

Figure~\ref{fig:dataset-one-pager-with-table} shows summary statistics and underlying distributions for each of the four source corpora and the VizNet 1M corpus.  The data type of a column is classified as either categorical, quantitative, or temporal, which we abbreviate as C, Q and T, respectively. This data type is detected using a heuristic-based approach that incorporates column name and value information. For quantitative columns, we use the Kolmogorov-Smirnov test ~\cite{massey1951kolmogorov} to examine the goodness-of-fit of six distributions: the normal, log-normal, exponential, power law, uniform and chi-squared distributions. We reject the null hypothesis of a distribution fit if the p-value of the associated test is lower than the level $\alpha=0.05$. If all distributions are rejected at $\alpha$, we consider the distribution to be undefined. If multiple distributions are not rejected, we consider the ``best" fit to be that with the highest p-value. We also report the skewness and percent of outliers, defined as data points that fall more than $1.5 \times IQR$ below the first quartile or above the third quartile, where $IQR$ is the interquartile range. The statistical distribution of categorical columns within each corpus is characterized using the normalized entropy.

\section{Experiment Design}

To evaluate the utility of VizNet as a resource for data scientists and visualization researchers, we conducted an experiment where we first replicated the Kim and Heer (2018) prior study~\cite{kim-and-heer} using real-world datasets from the VizNet corpus to assess the influence of user task and data distribution on visual encoding effectiveness. These datasets were sampled to match constraints from the prior study and ensure that participants only saw valid data. We then extended this experiment by including an additional task on outlier detection. Finally, we trained a machine learning model that learns the perceptual effectiveness of different visual designs and evaluated its predictive power across test datasets.



\begin{figure}[h]
    \centering
    \includegraphics[width=\columnwidth]{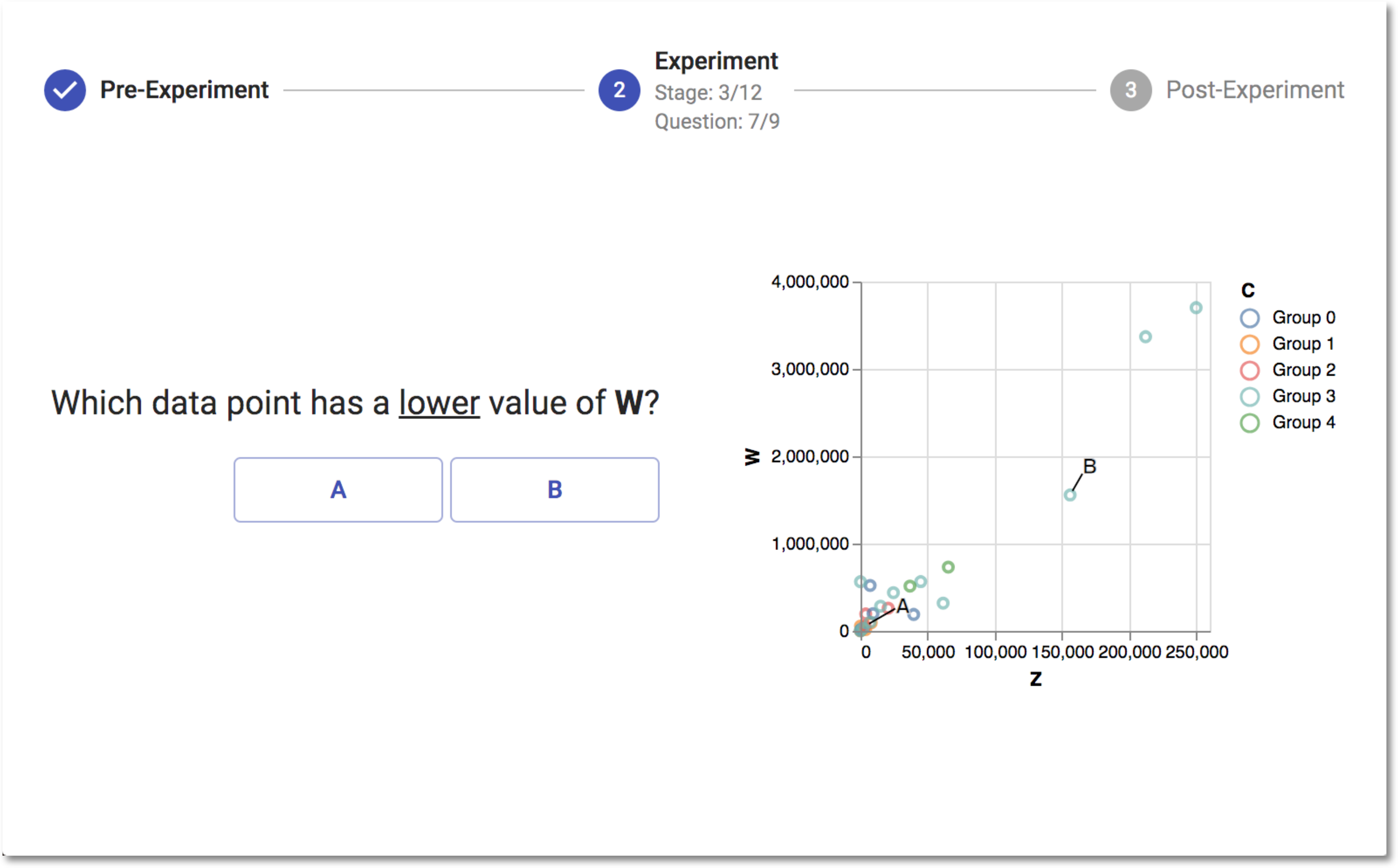}
    \caption{VizNet user interface for the \textit{Compare Values} task experiment.}
    \label{fig:screenshot}
\end{figure}

\subsection{Replication of Kim and Heer (2018)}
Kim and Heer (2018), \textit{``Assessing Effects of Task and Data Distribution on the Effectiveness of Visual Encodings,"}  conducted a crowdsourced experiment measuring subject performance (\textit{i.e.} error rate and response time) across data distributions (\textbf{D}), visualization designs (\textbf{V}), and task types (\textbf{T}). The 24 data distributions characterize trivariate data involving one categorical and two quantitative fields (C=1, Q=2) sampled from 2016 daily weather measurements~\cite{overview-of-the-global-historical} according to univariate entropies of the quantitative fields, cardinalities, and number of records per category. 

The authors employed a mixed design using a within-subjects treatment for visual encodings and between-subjects treatments for tasks and data characteristics. They analyzed responses from 1,920 participants on Amazon's Mechanical Turk (MTurk), who individually completed 96 questions and 12 engagement checks, and calculated the absolute and ranked performance of different ($\textbf{D} \times \textbf{V} \times \textbf{T}$) conditions, as well as the interaction effects between different data characteristics, visual channels, and task types. These results extended existing models of encoding effectiveness, such as APT~\cite{mackinlay:tog86}, and provided valuable insights for automatic visualization design systems.

\subsection{Datasets}
For this experiment, we sampled VizNet datasets according to a procedure that matched constraints from Kim and Heer (2018) and ensured that participants only saw valid data without missing values. This procedure was developed after an initial pilot study with a subset of the corpus in which all datasets were manually verified.

To begin, we identified all datasets with more than one categorical field and two quantitative fields (C$\geq$1 and Q$\geq$2). Then, we sampled all possible three column subsets with exactly one categorical and two quantitative fields (C=1, Q=2). Following this sampling, we filtered out datasets using a number of constraints. First, we rejected datasets containing any null values. Second, we required that the column names of all datasets must contain between 1 and 50 ASCII-encoded characters. Third, we limited the cardinality (\textit{e.g.} the number of unique groups) of the categorical columns between 3 and 30. Fourth, we restricted the group names between 3 and 30 characters, at least one of which is alphanumeric. Lastly, we required that each of the groups must contain 3 to 30 values. We chose these values to be consistent with the upper and lower constraints of Kim and Heer (2018).

Our sampling procedure resulted in 2,941 valid datasets from the Open Data Corpus (100,626 possible combinations), 6,090 valid datasets from Many Eyes (354,206 combinations), 1,368 from Plotly (347,387 combinations), and 82,150 from a subset of the Webtables corpus (1,512,966 combinations). From this set of candidates, we randomly selected 200 candidates per visualization specification $\times$ task condition. We use $\textbf{V}$ to denote the number of visualization specifications and \textbf{T} to denote the number of tasks, which leads to 60 such conditions ($\textbf{V} \times \textbf{T} = 12 \times 5 = 60$). The 200 number of datasets sampled from the VizNet corpus is consistent with the 192 datasets sampled in Kim and Heer (2018). As a result, this sampling resulted in $200 \times 12 = 2,400$ datasets per task, $2,400$ datasets per corpus, and $9,600 = 2,400 \times 4$ total datasets.

\subsection{Visual Encodings}
We selected the twelve visual encoding specifications chosen in Kim and Heer (2018). These encodings are specified using the Vega-Lite grammar~\cite{vegalite}, which specifies plots using a geometric mark type (\textit{e.g.} bar, line, point) and a mapping from data fields to visual encoding channels (\textit{e.g.} \textit{x}, \textit{y}, \textit{color}, \textit{shape}, and \textit{size}). In particular, Kim and Heer (2018) used twelve visualization designs, all of which are scatterplots (a \textit{point} mark) with different mappings between data and encoding channels.

We used the \textit{Tableau-10} scheme for color encoding categorical fields with cardinality less than 10, and \textit{Tableau-20} for categorical fields with cardinality greater than or equal to 20. For positional encodings, in contrast to Kim and Heer (2018), we used a heuristic to determine whether an axis should start at zero. If the range of a variable $Q$ is less than 10\% of maximum value $0.1 \times |\textit{max}(Q)|$, then we default to Vega-lite axis ranges. Based on a pilot study, we found that this heuristic was necessary to ensure that no questions were prohibitively difficult.

\subsection{Tasks}
Following Kim and Heer (2018), we considered 4 visualization tasks informed by the Amar et al. (2005) ~\cite{Amar2005} taxonomy of low-level analytic activities. Two of those tasks were \textit{value tasks}: \textit{Read Value} and \textit{Compare Values} asked users to read and compare individual values. The other two tasks were \textit{summary tasks}: \textit{Find Maximum} and \textit{Compare Averages} required the identification or comparison of aggregate properties. Each of these tasks was formulated as a binary question (two-alternative forced choice questions). We generated the two alternatives using the procedure described in the prior study.

\subsection{Procedure}
Identical to Kim and Heer (2018), we also employed a mixed design incorporating a within-subjects treatment for visual encodings and a between-subjects treatment for tasks. Each participant answered 9 questions (1 attention check and 8 real) for each of the 12 visual encodings, presented in a random order. Every participant was assigned to a specific task. Unlike Kim and Heer (2018), we did not incorporate dataset conditions. Each dataset was selected randomly from the pool of 200 datasets per \textbf{V} $\times$ \textbf{T} condition. In order to ensure reliable human judgment, we followed the process from Kim and Heer (2018) and incorporated 12 evenly distributed gold standard tasks. The gold standard tasks presented a user with a real dataset encoded in the present visual encoding condition, and asked what information is presented in the visual channel that encodes the first quantitative column ($Q_1$). 

\subsection{Participants}
Crowdsourcing platforms such as MTurk are widely used to recruit participants and conduct online experiments at scale~\cite{mason2012conducting, kittur2008crowdsourcing}. We recruited in total 1,342 MTurk workers who were located in the U.S. and had $\geq 95\%$ HIT approval rating.

During the analysis, we included the following criteria to ensure the quality of human judgment: we selected subjects who accurately answered 100\% of the gold standard questions, had an experimental error rate of less than 60\%, and can effectively distinguish colors. We had set the gold standard response exclusion threshold to 100\% (i.e., discarding responses if even 1 out of these 12 questions was answered incorrectly). We have verified that a more lenient 80\% exclusion threshold does not significantly change the results. Kim and Heer (2018) does not report a dropout rate, making it difficult to assess whether and by how much our dropout rate differs. We included two Ishihara color blindness plate tests ~\cite{ishihara1960tests} along with two pre-screen questions to ensure the participants can effectively distinguish colors. A total of 96.47\% reported no vision deficiency and were allowed to participate in the experiment. This resulted in a total of 624 participants' data for in the analysis. 

Of the 624 participants, 43.75\% were male, 55.44\% female, and 0.48\% non-binary. 6.38\% of the participants had no degree, whereas others had bachelor's (43.10\%), master's (14.90\%), Ph.D. (3.04\%), associate (14.58\%) degrees as well as a high school diploma (17.46\%). Each participant received 1.00 USD in compensation, which we calculated using the average times of a pilot study and the same hourly wage of Kim and Heer (2018). 




\section{Results}~\label{sec:results}
In this section, we describe the results of our experiment, compare them with the results of Kim and Heer (2018)~\cite{kim-and-heer}, and demonstrate a machine learning-based approach to predicting effectiveness from (\textit{data}, \textit{visualization}, \textit{task}) triplets.

\subsection{Comparing Subject Performance}
\begin{figure}[t]
    \centering
    \includegraphics[width=\columnwidth]{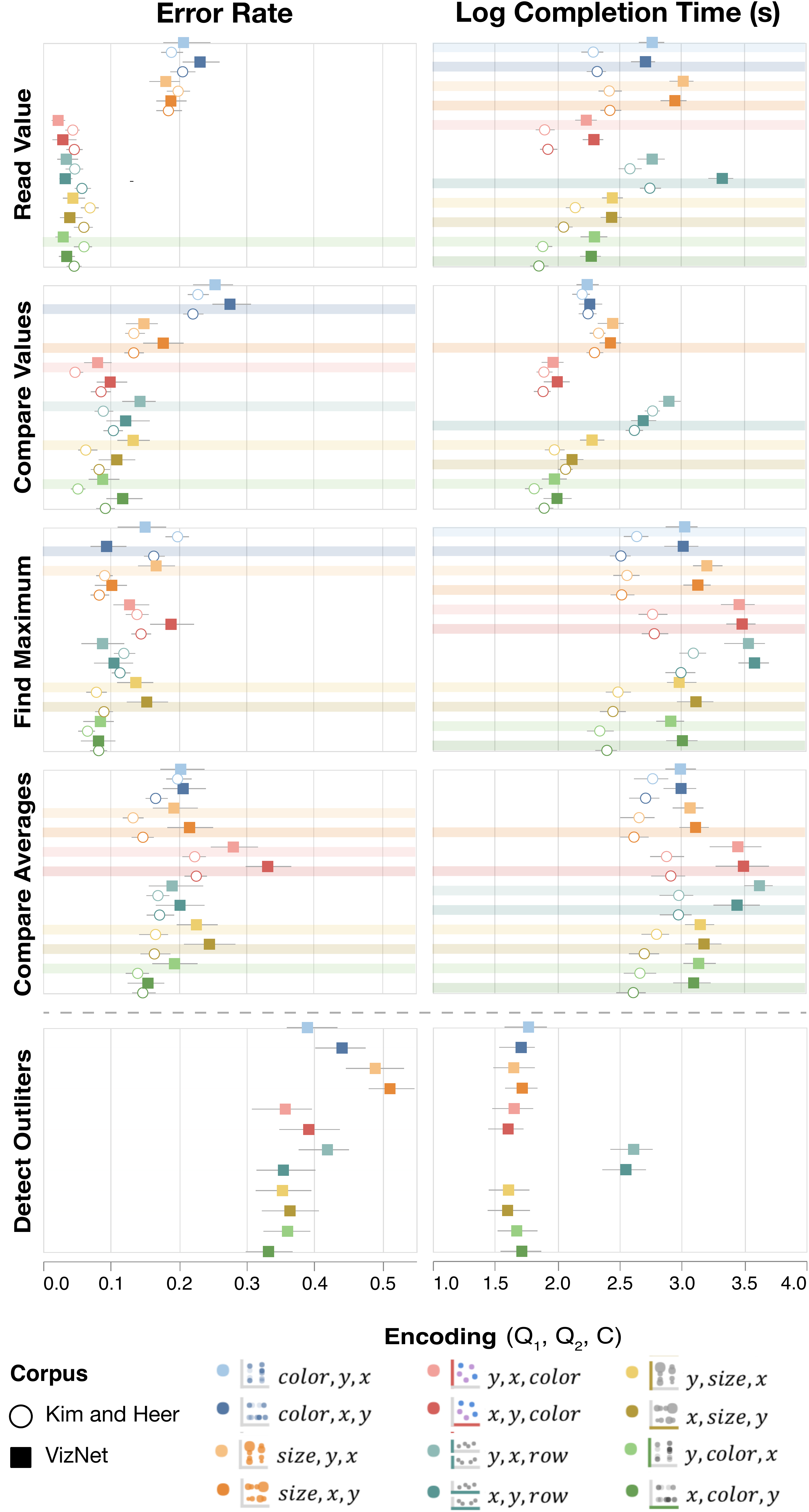}
    \caption{Bootstrapped means and 95\% confidence intervals for error rates (left) and log response times (right) across tasks and visual encodings for Kim and Heer (2018) original data, and our replication on VizNet. We reuse the original color encoding of Kim and Heer (2018). Shading indicates a statistically significant difference.}
    \label{fig:vt-error-rate-distributions}
\end{figure}

We first compared subject performance with the quantitative results of Kim and Heer (2018) by considering aggregate error rates and log response times per visualization specification and task condition (\textbf{V} $\times$ \textbf{T} = 12 $\times$ 4). Following this, we calculated mean error rates with 95\% bootstrapped confidence intervals, performed by sampling participants with replacement. To analyze the difference of mean error rates and response times we conducted permutation tests with $10^4$ permutations. We test significance at a significance level of $\alpha=0.05$ with Bonferroni correction for our $m=48$ hypotheses. The results for the error rate and log response times are shown in Figure~\ref{fig:vt-error-rate-distributions}.

The absolute error rates of our replication tend to agree with those of Kim and Heer (2018) for the \textit{Read Value} task, and to a lesser extent for the \textit{Compare Values} task. The rankings of different visual encodings are also similar. However, for the the \textit{summary tasks} (\textit{Find Maximum} and \textit{Compare Averages}), our observed error rates depart from those of Kim and Heer (2018). Though more data points are needed to draw meaningful conclusions, these results suggest that real-world data affects error rates for more complex tasks.

In contrast, the absolute response times in our study seem to be systematically longer for all tasks except the \textit{Compare Values} task. However, the relative rankings of different encoding are consistent with those of Kim and Heer (2018).

\subsection{Extending with an Outlier Detection Task}
As suggested by Kim and Heer (2018), investigating additional task types is a promising direction of future research. In particular, tasks with more subjective definitions, such as \textit{Cluster} and \textit{Find Anomalies} were not included in Kim and Heer (2018). Nevertheless, as outlier detection is one of the most important data analysis tasks in practice, it warrants further empirical study. We extended the prior work by considering this latter task of identifying ``which data cases in a set S of data cases have unexpected/exceptional values.''

We generated 2,400 datasets using the sampling methodology described in the previous section. First, we presented users with a definition of outliers as ``observations that lie outside the overall pattern of distribution.'' Then, using the same experiment design, we assessed answers to the question ``Are there outliers in $Q_1$?'' ``Yes'' and ``No'' are provided as response options. Outliers were determined using the median absolute deviation (MAD)-based approach described in~\cite{outliers}, which is robust to varying sample sizes, compared to other simple approaches.

We found that the error rates for the outlier detection task are higher compared to the other tasks (see Figure~\ref{fig:vt-error-rate-distributions}). This may be due to an inadequate measure of ground truth, inconsistent definitions, or lack of prior training. It is important to note that the specification rankings resemble that of the \textit{Read Value} task: \emph{color} and \emph{size} trail behind other encodings channels. Conversely, the log response times are significantly shorter than for other tasks, for all except the faceted charts with \emph{row} encodings.

\subsection{Learning a Model to Predict Effectiveness}
To characterize a dataset, we extracted 167 features: 60 per quantitative field Q, 11 for the categorical field C, 15 for the Q-Q pair, 6 for the two C-Q pairs, and 9 which consider all three fields. These features characterized summary statistics (\textit{e.g.} coefficient of variance and kurtosis), statistical distributions (\textit{e.g.} entropy and statistical fits), pairwise relationships (\textit{e.g.} correlations and one-way ANOVA p-values), clusteredness and spatial autocorrelation.

\begin{figure}[t]
    \centering
    \includegraphics[width=\columnwidth]{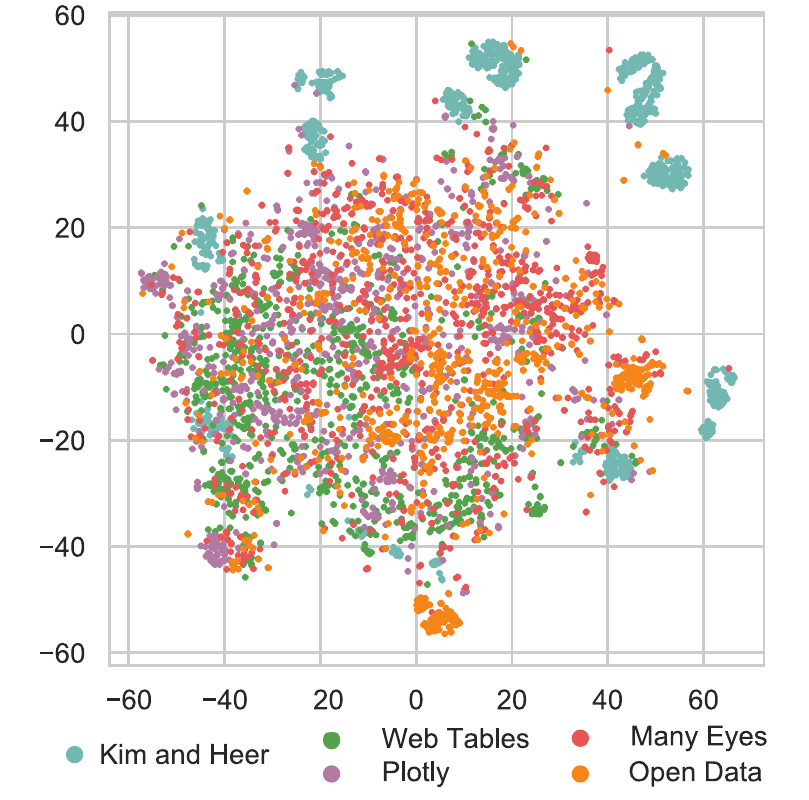}
    \caption{Two-dimensional t-SNE projection of datasets with one categorical and two quantitative columns, evenly sampled from Kim and Heer (2018) and the four corpora within VizNet, with a perplexity of 75.}
    \label{fig:tsne}
\end{figure}

We first decoded diversity within our space of datasets using these features. Using principal components analysis, we computed 32 principal components which collectively explain over 85\% of the variance within our dataset. Then, we generated a two-dimensional t-SNE projection of these principal components, as shown in Figure~\ref{fig:tsne}. It is important to note that the datasets used in Kim and Heer (2018) ~\cite{kim-and-heer} are highly clustered and separate from the datasets used within our replication. This observation is robust for different numbers of principal components and values of perplexity (5-200).

\begin{figure}[t]
    \includegraphics[width=0.9\columnwidth]{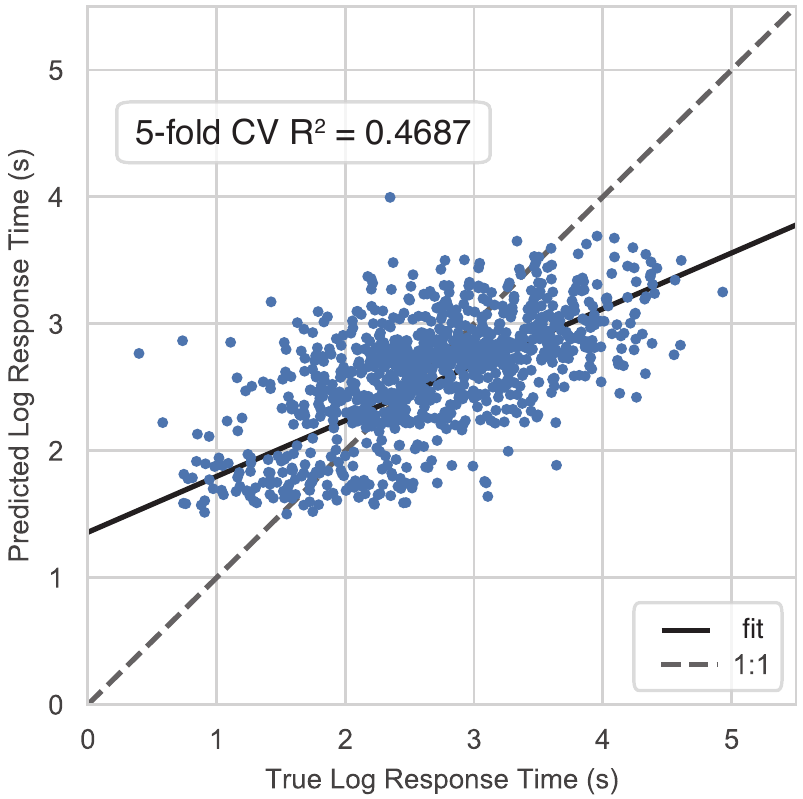}
    \caption{Observed log response times (in seconds) vs. those predicted by a gradient boosted regression tree. The dotted diagonal line denotes a perfect relationship between observation and prediction.}
    \label{observed-vs-predicted-response-times}
\end{figure}

\begin{figure}[t]
    \centering
    \includegraphics[width=\columnwidth]{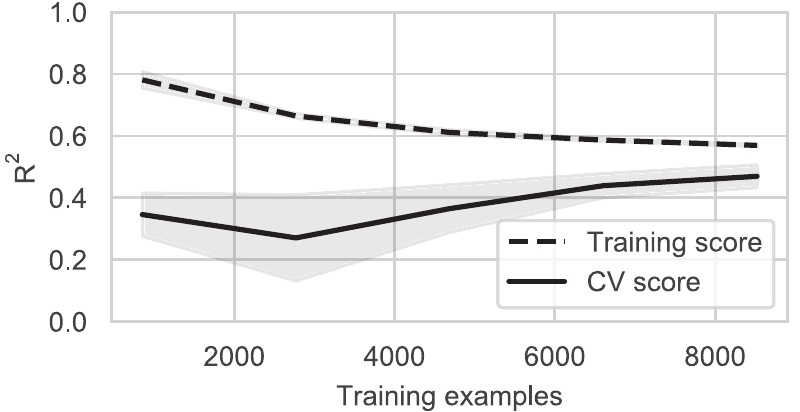}
    \caption{Training $R^2$ and 5-fold cross-validation $R^2$ as the number of training examples increases.}
    \label{response-time-prediction-learning-curves}
\end{figure}

To predict log completion time we use gradient boosted regression trees, a model with strong ``off-the-shelf'' performance. Training on 80\% sample of the data, we were able to predict log completion times in a 20\% hold-out test set with a 5-fold cross-validated $R^2$ of 0.47, which strongly outperforms baseline models such as K-nearest neighbors and simple linear regression. A scatter plot of observed vs. predicted values for the top performing model is shown in Figure~\ref{observed-vs-predicted-response-times}. Learning curves in Figure~\ref{response-time-prediction-learning-curves} indicate that, despite the large number of features, our model does not overfit on the training set, and that there are still gains from increasing the number of training samples.

Kim and Heer (2018) reports the trade-off between response time and error rate. To capture this trade-off, we created a combined metric from the log response times and error rate metrics by partitioning the log response times into 20\% quantiles, and the error rates into five bins of equal width, for a total of 25 pairs. Then, we characterized each \emph{(d,\,v,\,t)} triplet with the associated (response time + error rate) pair, and resampled minority classes using the Synthetic Minority Over-sampling Technique (SMOTE)~\cite{smote}. Training a gradient boosted classification tree on the balanced training set resulted in a Top-3 prediction accuracy of 52.48\%.  

\subsection{Limitations}
Although we have successfully demonstrated the effectiveness of VizNet, it is important to acknowledge limitations. Replication and reproducibility are essential to advance research~\cite{peng2011reproducible}. In the experiment, we attempted to replicate Kim and Heer (2018) as closely as possible. However, due to practical constraints, we introduced clarifying modifications to the question text and interface design. Due to variance between crowd workers, we were not able to recruit the same participants; nor do we control for question difficulty, which is calibrated in Kim and Heer (2018). Most of all, we did not exactly replicate the original conditions of the synthetic datasets, which would have limited the amount of real-world VizNet datasets available for sampling. Notwithstanding these limitations, our work provides an important direction to understand the opportunities and challenges faced in replicating prior work in human-computer interaction and visualization research. 

With respect to extending the experiment to include an additional task, we note that outlier detection, unlike the other tasks, does not have a defined ground truth. Though we used a robust outlier detection method, there may be a limitation to any purely quantitative method that does not rely on human consensus. The lack of an objective notion of outliers and absence of a clear definition thereof in the questions, reinforces the inconsistency between ground truth and crowdsourced labels presumably partially explaining the consistently high error rate. In the context of the machine learning model, while human judgments can play an important role in help predicting perceptual effectiveness, crowdsourced training data can be noisy. The current experiment was unable to analyze lower bound requirements of quality data, but VizNet's diverse dataset offers such opportunity for future research.

\section{Discussion} 
There are several important areas where VizNet makes important contributions. VizNet provides a noteworthy contribution to advance our knowledge of effective graphical perception by enabling scientific community access to rich datasets for visualization learning, experimentation, replication, and benchmarking. VizNet offers both the full corpus and the sampled corpus of one million datasets (VizNet 1M). It further described the dimensions, types, and statistical properties of these datasets. The voluminous collection of VizNet complements synthetically generated data. Moreover, the properties of the VizNet corpus can inform assessments of the ecological validity of other corpora from domains beyond VizNet. 

\textbf{Implications of enabling the VizNet interface for the scientific community.} We envision that in the long run, adoption of a common corpus and benchmarks by the visualization community will facilitate the sharing and comparing of results at scale. We have made VizNet publicly available at \href{https://viznet.media.mit.edu}{\texttt{https://viznet.media.mit.edu}}. A taxonomy in VizNet is formed by splitting our corpus first on the number of columns of a datasets, and then on the composition of column types. Therefore, we should design interactions to help users query, filter, sample datasets within this taxonomy (\textit{e.g.} give me all datasets with one categorical, two quantitative, and one temporal field). Moreover, this informs the need for supporting keyword search to allow filtering by domain, in addition to filtering on other dataset properties (\textit{e.g.} give me highly correlated datasets with exactly two quantitative fields).


\textbf{Implications of VizNet for replication and experimentation.} We replicate Kim and Heer (2018) to demonstrate the utility of using VizNet. Our results with real-world data are largely consistent with their findings. As a result of our more diverse backing datasets, however, there are statistically significant differences in error rates for the complex tasks. We also note that task completion times with real data are consistently longer for all but one task. These discrepancies suggest that graphical perception studies must account for the variation found in real datasets. Kim and Heer (2018) acknowledge this direction of future work by describing the need for investigating ``all [data] distributions of potential interest.'' The process of harvesting these diverse distributions would be facilitated by using VizNet. We further extend the original experiment by considering an additional ``detect outliers'' task, an important but subjective visual analysis task that is difficult to assess using synthetic data.

\textbf{Implications of VizNet for learning a metric of perceptual effectiveness.} While Kim and Heer (2018) employed a mixed effects model to analyze their results, we proposed to conceive the harvested data as a collection of \emph{(data, visualization, task)} triplets, each of which is associated with effectiveness measures. Using machine learning models, we predicted the completion time with an $R^2$ value of 0.47. Acknowledging the trade-off between completion time and error rate, we constructed a combined metric and achieved a top-3 prediction accuracy of 52.48\%. Despite the noise and skew of crowdsourced labels, and a relatively small sample size, these results out-perform both random chance and baseline classifiers. In doing so, they illustrate the potential for learning a metric of perceptual effectiveness from experimental results.

\section{Future Work} We plan to extend VizNet along three major directions: (1) incorporate and characterize more datasets, (2) harness the wisdom of the crowd, and (3) develop active learning algorithms for optimal experiment design. 

\textbf{Incorporate and characterize more datasets.} VizNet currently centralizes four corpora of data from the web, open data portals, and online visualization galleries. We plan to expand the VizNet corpus with the 410,554 Microsoft Excel workbook files (1,181,530 sheets)~\cite{Chen:2017:SPD:3132847.3132882} extracted from the ClueWeb09 web crawl\footnote{\url{https://lemurproject.org/clueweb09.php}}. Furthermore, Morton et. al.~\cite{morton-public-data-and-visualizations} report $73,000$ Tableau workbooks and $107,500$ datasets from Tableau Public, which could be integrated into VizNet. Lastly, we plan to incorporate $10,663$ datasets from Kaggle\footnote{\url{https://kaggle.com/datasets}}, $1,161$ datasets included alongside the R statistical environment\footnote{\url{https://github.com/vincentarelbundock/Rdatasets}}, and to leverage the Google Dataset Search\footnote{\url{https://toolbox.google.com/datasetsearch}} to source more open datasets.


In the future work, we plan to characterize the semantic content within column and group names using natural language processing techniques such as language detection, named entity recognition, and word embeddings. Moreover, as we describe the features of datasets within the VizNet corpus, we can characterize the bias between corpora in terms of dimensions, type composition, and statistical properties of columns. This will enable us to systematically study the extent to which these corpora differ. The existence of such bias between corpora is clear from the previous data section ~\cref{sec:data}. A clearer understanding of between-corpus bias could inform future techniques for sampling from the VizNet corpus. 

\textbf{Harness the wisdom of the crowd.} Domain specific crowdsourcing platforms such as FoldIt, EteRNA, GalaxyZoo, and Game with Purpose, have incentivized citizen scientists to discover new forms of proteins~\cite{cooper2010predicting}, RNAs~\cite{lee2014rna}, galaxies~\cite{lintott2008galaxy}, and artificial intelligence algorithms~\cite{von2008designing}. We envision VizNet will enable citizen scientists and visualization researchers to execute graphical perception experiments at scale. In recent years, crowdsourcing has been pivotal in the creation of large-scale machine learning corpora. Daemo~\cite{gaikwad2015daemo}, a self-governed crowdsourcing marketplace, was instrumental in the creation of the Stanford Question Answering Dataset (SQuAD)~\cite{Rajpurkar2016-SQuAD}, whereas MTurk was used to curate the ImageNet dataset\cite{Deng09imagenet:a}. 

The effectiveness of the crowdsourcing has also been exemplified in our experiment while collecting the human judgments for the critical evaluation of visual designs. It is interesting to note that some of the crowd workers enjoyed the intellectual aspect of the experiment, as illustrated by post experiment responses: (1) \textit{`I found this survey entertaining, it makes you think and use your head'} (2) \textit{`It is a very interesting survey to carry out since it promotes the capacity of analysis I congratulate you for that'}. A natural progression to harness crowdsourcing mechanisms for VizNet includes extension of literature on task design ~\cite{kittur2013future}, crowd work quality improvements ~\cite{le2010ensuring, dow2012shepherding}, and incentive design ~\cite{gaikwad2016boomerang,von2008designing}.   

\begin{figure}[t]
    \centering
    \includegraphics[width=\columnwidth]{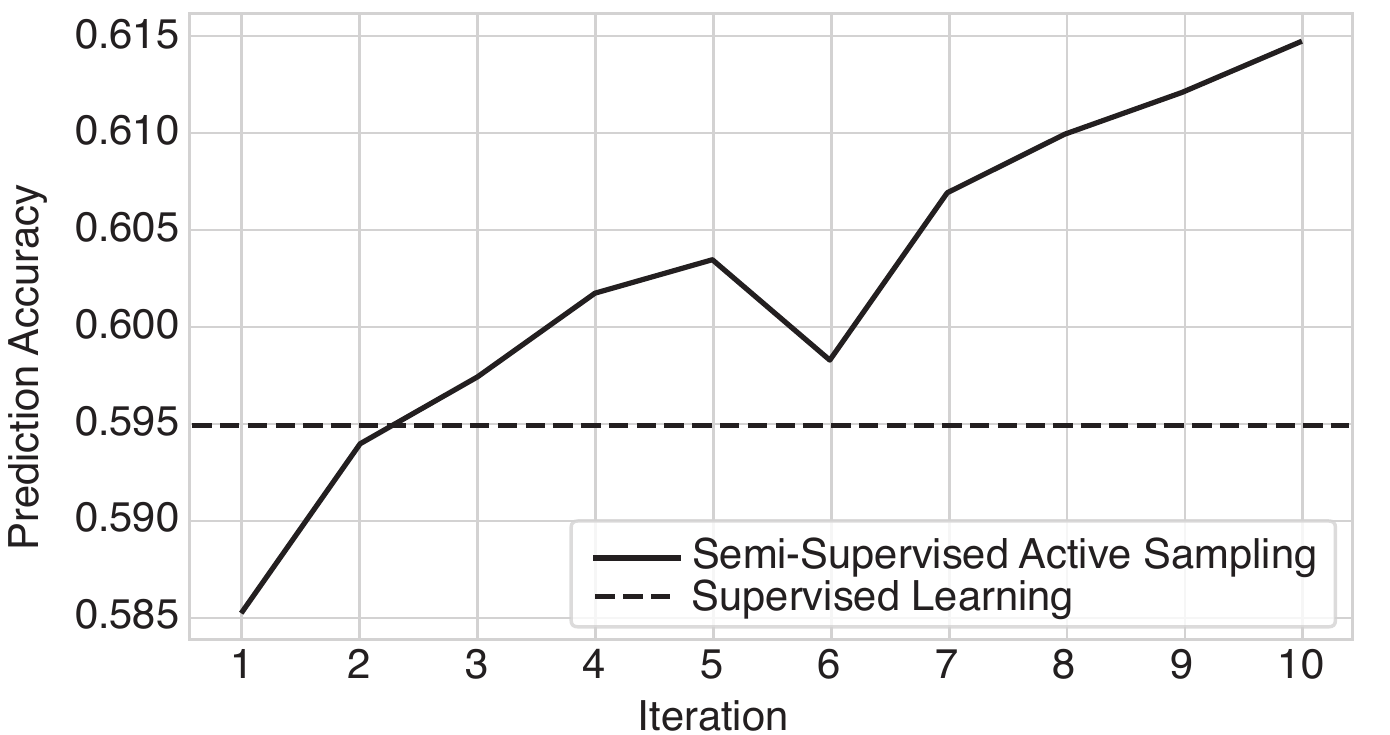}
    \caption{Performance curves obtained by semi-supervised active learning and supervised learning over 10 iterations.}
    \label{fig:active-learning-curve}
\end{figure}

\textbf{Develop active learning for optimal experiment design.} Although gathering human-judgment labels for each triplet is costly, it is possible to learn the effectiveness from labeled triplets to predict labels for unseen ones (see section ~\cref{sec:results}). In order to further illustrate this strategy we conducted a small experiment on the same data as in section ~\cref{sec:results} where the completion times are categorized into low, medium and high. To propagate labels we employed self-learning ~\cite{agrawala1970learning}, so we added the model predictions with high certainty to the labelled set. The predictions with low certainty were replaced with crowdsourced labels following the uncertainty algorithm ~\cite{cohn1996active}. Figure \ref{fig:active-learning-curve} shows how this strategy improves the accuracy on a test set after a number of iterations against the baseline of training on all labeled samples (supervised learning). In the future, we plan to harness active learning to assess the quality of human judgment.


\section{Conclusion}
Large-scale data collection efforts for facilitating research are common across sciences and engineering, from genomics to machine learning. Their success in accelerating the impact of research in respective fields is a testament to the importance of easy access to large-scale realistic data as well as benchmarking and performing research on shared databases. As the field of data visualization research grows from its infancy, we expect the need for and utility of large-scale data and visualization repositories to significantly grow as well. VizNet is a step forward in addressing this need.

\bibliographystyle{ACM-Reference-Format}
\balance
\bibliography{bibliography,graphical-perception}

\end{document}